\documentclass[prb,preprint,citeautoscript]{revtex4-1}
\usepackage[breaklinks,colorlinks]{hyperref}
\usepackage{graphicx}
\usepackage{color}
\usepackage{amsmath}
\usepackage{amssymb}
\usepackage{amstext}
\usepackage{bm}
\usepackage{breakurl}
\usepackage{tensor}
\usepackage[utf8]{inputenc}

\newcommand{\beq}{\begin{equation}}
\newcommand{\eeq}{\end{equation}}
\newcommand{\bx}{\mathbf{x}}
\newcommand{\bv}{\mathbf{v}}
\newcommand{\bnab}{\boldsymbol{\nabla}}
\newcommand{\nn}{\nonumber}

\begin{document}

\title{Waves and null congruences in a draining bathtub}

\author{David Dempsey}
\email{ddempsey1@sheffield.ac.uk}

\author{Sam R.~Dolan}
\email{s.dolan@sheffield.ac.uk}

\affiliation{Consortium for Fundamental Physics, School of Mathematics and Statistics,
The University of Sheffield, Hicks Building, Hounsfield Road, Sheffield S3 7RH, United Kingdom.}

\date{\today}

\begin{abstract}
We study wave propagation in a draining bathtub: a fluid-mechanical black hole analogue in which perturbations are governed by a Klein-Gordon equation on an effective Lorentzian geometry. Like the Kerr spacetime, the draining bathtub geometry possesses an (effective) horizon, an ergosphere and null circular orbits. We propose that a `pulse' disturbance may be used to map out the light-cone of the effective geometry. First, we apply the eikonal approximation to elucidate the link between wavefronts, null geodesic congruences and the Raychaudhuri equation. Next, we solve the wave equation numerically in the time domain using the method of lines. Starting with Gaussian initial data, we demonstrate that a pulse will propagate along a null congruence and thus trace out the light-cone of the effective geometry. Our numerical results reveal features, such as wavefront intersections, frame-dragging, winding and interference effects, that are closely associated with the presence of null circular orbits and the ergosphere.
\end{abstract}


\maketitle

\section{Introduction}

The first direct observation of gravitational waves GW150914 undoubtedly heralds a new era in astronomy\cite{Abbott:2016blz}. The ``chirp'', observed in both Advanced LIGO (aLIGO) detectors, bears the widely-anticipated imprint of a binary black hole merger\cite{TheLIGOScientific:2016wfe}. The beautiful match between the observed signal and the `model' waveform surely dispels any lingering doubts about the reality of Kerr-like black hole solutions in nature, and the efficacy of Einstein's theory of general relativity in the strong-field regime. In 1975, Chandrasekhar\cite{Chandrasekhar:1987} reflected that ``the most shattering experience has been the realization that [Kerr's] solution of Einstein's equations of general relativity provides the absolutely exact representation of untold numbers of massive black holes that populate the universe.'' With between 3 and 90 significant black hole merger detections expected in the next observing run at aLIGO\footnote{The First Measurement of a Black Hole Merger and What it Means: \url{http://www.ligo.org/science/Publication-GW150914Astro/index.php}}, we may share in this prescient revelation. 

While we delight in this new window on astrophysical black holes, a complementary effort is underway to observe black hole \emph{analogues} in a laboratory setting. In 1981, Unruh \cite{Unruh:1981} noted ``the model of the behaviour of a quantum field in a classical gravitational field is the motion of sound waves in a convergent fluid flow''.  In 2010, Unruh's team reported on a measurement of stimulated Hawking emission in a wavetank experiment\cite{Weinfurtner:2010nu, Weinfurtner:2013}. The key insight underpinning fluid-mechanical analogues\cite{Unruh:1981, Visser1998} is that, under certain assumptions\footnote{assuming that a barotropic, inviscid fluid undergoes a stable irrotational flow\cite{Visser1998}}, the linearized equation governing an acoustic disturbance in the velocity potential of a fluid flow is formally equivalent to the d'Alembertian equation for a scalar field $\psi$ propagating in a (3+1)-dimensional Lorentzian geometry, 
\beq \label{eq:wave}
\Box \psi \equiv \frac{1}{\sqrt{|g|}} \partial_\mu(\sqrt{|g|} g^{\mu \nu} \partial_\nu \psi) =0.
\eeq
Here, the fluid velocity is $\bv = \bv_0 + \delta \bv$ where $\bv_0$ is the background flow velocity and $\delta \bv = - \bnab \psi$ is a small perturbation. In essence, perturbations in the background flow propagate as if experiencing an \emph{effective geometry} described by the line element
\beq
ds^2 \equiv g_{\mu \nu} dx^\mu dx^\nu = \frac{\rho}{c} \left\{ -c^2 dt^2 + (d\bx - \bv_0 dt) \cdot  (d\bx - \bv_0 dt)\right\} . \label{eq:lineelement}
\eeq
The effective metric $g_{\mu \nu}$ depends only on the background flow parameters -- velocity $\bv_0(\bx, t)$, density $\rho$ and speed of sound $c$ -- rather than on Einstein's equations. Thus, an ingenious experimenter may, in effect, sculpt their own spacetime. 

Various alternative analogue systems have been proposed, in acoustics, optical materials, and condensed matter theory \cite{Novello:2002qg, Barcelo:2005fc, Rousseaux:2010md}. Sch\"utzhold and Unruh\cite{Schutzhold:2002} suggested using surface waves on a fluid in a shallow basin. They showed that the velocity potential for long-wavelength gravity waves is also governed by Eq.~(\ref{eq:wave}), and the effective geometry takes the form (\ref{eq:lineelement}). They proposed a two-dimensional (surface wave) version of the \emph{draining bathtub} model introduced by Visser\cite{Visser1998}, in which the background velocity is 
\beq
\bv_0 = \frac{1}{r} \left(C \hat{\boldsymbol{\phi}} - D \hat{\mathbf{r}}\right) , \label{eq:velocity}
\eeq 
and $C$ and $D$ are circulation and draining constants, respectively. This flow is divergence-free and irrotational ($\bnab \cdot \bv = 0 = \bnab \times \bv$), except at the sink. The flow possesses an \emph{effective horizon} at $r_h = D/c$, where the radial flow speed exceeds the speed of sound. It also possesses an ergosphere within $r_e = \sqrt{C^2 + D^2}/c$, where all perturbations are co-rotating with the background flow. Henceforth we assume $\rho$ and $c$ are constant, and adopt units in which they are equal to unity. 

A range of aspects of the draining bathtub model have been investigated: superradiance \cite{Basak:2002aw, Berti:2004ju}, absorption \cite{Oliveira:2010zzb}, quasinormal ringing \cite{Cardoso:2004fi, Dolan:2011ti}, wave scattering \cite{Dolan:2012yc} and the (modified) Aharonov-Bohm effect \cite{Dolan:2011zza}. 

Here we address a simple question: if a small `pulse' disturbance is created at a point in the draining bathtub flow (for instance, by dropping a stone), how does the resulting wave propagate? We shall show via approximate and numerical methods that, in essence, the wavefront emanating from the original disturbance maps out the `light-cone' of the effective geometry. We hope that this possibility will be explored in wavetank experiments in the near future\cite{QGL}. 

This report proceeds as follows: In Sec.~\ref{sec:method} we outline methods for solving Eq.~(\ref{eq:wave}); in Sec.~\ref{sec:results} we present sample results; and in Sec.~\ref{sec:conclusions} we discuss and conclude. Throughout, we use a positive metric signature, and cylindrical coordinates $\{t, r, \phi\}$ tied to the laboratory frame.  Greek letters $\mu, \nu, \ldots$ denote indices in the effective spacetime. Indices are lowered and raised with the effective metric $g_{\mu \nu}$ and its inverse $g^{\mu \nu}$. Partial derivatives are denoted with commas, and covariant derivatives with semi-colons. 

\section{Method\label{sec:method}}
In Sec.~\ref{sec:eikonal}, we review the eikonal approximation which reveals the relationship between short-wavelength perturbations and null geodesic congruences. Next, we describe methods for solving the geodesic equations (Sec.~\ref{sec:geodesic}), and the transport equations for (e.g.) the expansion scalar along a congruence (Sec.~\ref{sec:transport}). Finally, we outline our method for solving the wave equation itself (Sec.~\ref{sec:waveeq}). The results of these approaches are compared in Sec.~\ref{sec:results}. 

 \subsection{The eikonal approximation\label{sec:eikonal}}
Let us start by considering wave propagation on a $(d+1)$-dimensional Lorentzian manifold (where $d=3$ in spacetime, and $d=2$ in the draining bathtub model). We consider a short-wavelength perturbation $\psi$, governed by Eq.~(\ref{eq:wave}), whose phase varies rapidly over a much shorter scale than $\sqrt{\mathcal{R}}$, where $\mathcal{R}$ is the typical curvature scale of the geometry\cite{Hollowood:2008kq}. We introduce the {\it ansatz},
\beq
\psi(x) = \mathcal{A}(x) \exp\left( i \omega \Theta(x) \right) , \label{eq:eikonal}
\eeq
where $\mathcal{A}(x)$ and $\Theta(x)$ are (eikonal) amplitude and phase functions, and $\omega$ is an order-counting parameter. We proceed by inserting (\ref{eq:eikonal}) into (\ref{eq:wave}) and expanding order-by-order in $\omega$. At leading order, $O(\omega^2)$, is the eikonal equation
\beq
g^{\mu \nu} \Theta_{, \mu} \Theta_{, \nu} = 0 . \label{eq:null}
\eeq
The gradient of the eikonal phase, $k^\mu \equiv \Theta^{,\mu}$, is a null vector field that is normal to the family of constant-phase hypersurfaces $\Theta(x) = \text{const}$. 
By taking a derivative of Eq.~(\ref{eq:null}) and using the identity $k_{\mu ; \nu} = k_{\nu ; \mu}$ (as $k_\mu$ is a gradient) it is straightforward to show that $k^\mu$ satisfies the equation of parallel transport, $k^\mu k_{\nu ; \mu} = 0$. Thus, the integral curves of $k^\mu$ are null geodesics. In other words, the vector field $k^\mu$ defines a \emph{null congruence}\cite{Poisson:2004relativist}, i.e., a family of null geodesics whose tangent vectors are given by $k^\mu$. 
Thus, the eikonal wavefront propagates along a (hypersurface-orthogonal) null congruence.

At next-to-leading-order $O(\omega^1)$ one obtains a transport equation for the eikonal amplitude,
$
k^\mu \mathcal{A}_{,\mu} = -\frac{1}{2} \tensor{k}{^\nu _{;\nu}},  
$
which may be written as 
\beq
\frac{d \mathcal{A}}{d \lambda} = - \frac{1}{2} \theta,  \label{eq:amplitude}
\eeq
where $\theta \equiv \tensor{k}{^\nu _{;\nu}}$ is the \emph{expansion scalar} for a null congruence, and the derivative with respect to the affine parameter $\lambda$ is defined along a null geodesic, that is, an integral curve of $k^\mu$. The expansion scalar $\theta$ describes the rate at which the crosssection of the geodesic congruence expands ($\theta > 0$) or shrinks ($\theta < 0$).

The focussing of neighbouring null geodesics is described by the Raychaudhuri equation\cite{Poisson:2004relativist},
\beq \label{eq:rayc}
\frac{d\theta}{d \lambda} = - \frac{\theta}{d-1} - \sigma^{\mu \nu} \sigma_{\mu \nu} + \omega^{\mu \nu} \omega_{\mu \nu} - R_{\mu \nu} k^\mu k^\nu,
\eeq
where $\sigma_{\mu \nu}$ is the shear tensor, $\omega \equiv k_{[\mu ; \nu]}$ is the vorticity tensor, and $R_{\mu \nu}$ is the Ricci tensor for the effective spacetime. In the eikonal case, the null congruence is hypersurface-orthogonal and thus the vorticity is zero, $\omega_{\mu \nu} = 0$. Furthermore, in the case $d=2$, the shear tensor for null congruences is identically zero, and the Raychaudhuri equation reduces to
\beq \label{eq:rayc2}
\frac{d\theta}{d \lambda} = - \theta - R_{\mu \nu} k^\mu k^\nu .
\eeq

 \subsection{Geodesic equations\label{sec:geodesic}}
A geodesic is a path $q^\mu(\lambda)$ that extremizes the action functional $S[q^\mu(\lambda)] = \int L(q^\mu, \dot{q}^\mu) d \lambda$ with the Lagrangian $L(q^\mu, \dot{q}^\mu) \equiv \frac{1}{2} g_{\mu \nu} \dot{q}^{\mu} \dot{q}^{\nu}$ where $\dot{q}^\mu = dq^\mu / d\lambda$. The canonical momentum $p_\mu$ is defined by $p_\mu \equiv \frac{\partial L}{\partial \dot{q}^\mu} = g_{\mu \nu} \dot{q}^\nu (= k_\mu)$, and the corresponding Hamiltonian is $H[q^\mu, p_\nu] = \frac{1}{2} g^{\mu \nu} p_\mu p_\nu$. Thus, the geodesics may be found by solving Hamilton's equations, $\dot{q}^\mu = \frac{\partial H}{\partial p_\mu}$ and $\dot{p}_\mu = -\frac{\partial H}{\partial q^\mu}$. 

In the draining bathtub case, $p_t$ and $p_\phi$ are constants of motion, since $g^{\mu \nu}$ (and thus $H$) does not depend on $t$ or $\phi$. We may set $p_t = -1$ without loss of generality, as rescaling $p_t$ is equivalent to rescaling the affine parameter $\lambda$. Furthermore, $H$ is a constant of motion, and $H=0$ for null geodesics. This leads to an energy equation\cite{Dolan:2012yc},
\beq
\dot{r}^2 = U(r; p_\phi) , \quad \quad U(r; p_\phi) \equiv  \left(1 - \frac{C p_\phi}{r^2} \right)^2 - \left(1 - \frac{D^2}{r^2} \right) p_\phi^2 .
\eeq
The conditions $U = 0 = \partial_r U$ may be solved to locate the co-rotating ($+$) and counter-rotating ($-$) null circular orbits at radii $r_c^{\pm}$, where
\begin{align}
p_\phi^\pm &= \pm 2 \sqrt{C^2 + D^2} - 2 C , \nn \\
r_c^\pm &= \left( \sqrt{C^2 + D^2} |p_\phi^\pm| \right)^{1/2} . \label{eq:rc}
\end{align}

In practice, we found individual null geodesics emanating from a point by solving the Euler-Lagrange equations (rather than Hamilton's equations). To track a `spray' of geodesics within a congruence, we naturally wish to use the laboratory time $t$ rather than the affine parameter $\lambda$. A simple practical solution was to apply the chain rule $\frac{d}{d \lambda} = \dot{t} \frac{d}{dt}$ to convert to equations with $t$ as the independent parameter.

\subsection{Transport equations\label{sec:transport}}
To find the amplitude $\mathcal{A}(x)$ on a wavefront, we may evolve the transport equations (\ref{eq:amplitude}) and (\ref{eq:rayc2}) along each geodesic in a spray representing a congruence. As we are considering here a point-like disturbance, we wish to consider all null geodesics emanating from a spacetime event at $\lambda = 0$. A practical issue arises: the expansion scalar $\theta$ (and thus the amplitude $\mathcal{A}$) diverges as $\theta \sim (d-1)/\lambda$ in the limit $\lambda \rightarrow 0$. To handle this more gracefully, we may introduce the van Vleck determinant $\Delta$ (see e.g.~Ref.~\cite{Visser:1992pz}), which approaches unity in this limit, and which is governed by
\begin{equation} \label{eq:vvdet}
\frac{d \Delta}{d \lambda} = \left(\frac{d-1}{\lambda} - \theta \right) \Delta.
\end{equation}
By comparing the transport equations Eq.~(\ref{eq:vvdet}) and Eq.~(\ref{eq:amplitude}), it is straightforward to establish that the eikonal amplitude is related to the van Vleck determinant as follows,
\beq
\mathcal{A} = \frac{\sqrt{\Delta}}{\lambda^{(d-1)/2}} .  \label{eq:Avv}
\eeq
Thus, the wavefront's amplitude is found by computing the van Vleck determinant along a sample of representative geodesics in a null congruence.

 \subsection{Numerical solution of the wave equation\label{sec:waveeq}}
We begin by decomposing $\psi$ into azimuthal modes using the {\it ansatz}
\beq
\psi(t,r,\phi) = \sum\limits_{m=-\infty}^\infty \psi_m(t,r)  \frac{e^{im\phi} }{\sqrt{r}}, \label{eq:mmode}
\eeq
and noting the reality condition $\psi_{-m} = \psi^*_m$. Inserting (\ref{eq:mmode}) into (\ref{eq:wave}) with the effective metric (\ref{eq:lineelement}) and flow profile (\ref{eq:velocity}) leads to a set of 1+1D partial differential equations
\begin{align}
- \frac{\partial^2 \psi_m}{\partial t^2}
+ \frac{2D}{r}  \frac{\partial^2 \psi_m}{\partial t \partial r}
+ \left(1 - \frac{D^2}{r^2} \right) \frac{\partial^2 \psi_m}{\partial r^2} \quad \quad \nn \\
- \frac{ (D+2 i m C)}{r^2} \frac{\partial \psi_m}{\partial t} + \frac{2 D(D+i m C)}{r^3} \frac{\partial \psi_m}{\partial r} \quad \quad & \nn \\
- \left(\frac{\left(m^2-1/4\right)}{r^2} + \frac{- m^2 C^2 + 3 i m C D + 5 D^2 / 4}{r^4} \right) \psi_m
&=0. \label{eq:1+1}
\end{align}

\subsubsection{Initial data\label{sec:initialdata}}
We take as our initial data a Gaussian `pulse' in the field: $\psi(0, \bx) = \psi_0(r,\phi)$, $\partial_t \psi(0, \bx) = 0$, with
\beq
\psi_0(r,\phi) = \frac{1}{2 \pi \sigma^2} \exp\left(- \frac{(r\cos\phi-x_0)^2 + (r\sin\phi - y_0)^2}{2 \sigma^2} \right).
\eeq
Here $(x_0, y_0)$ are the Cartesian coordinates of the centre of a Gaussian pulse of width $\sigma$. Below we take $x_0 = -10$, $y_0 = 0$ and $\sigma = 1$ (and $C = D = 1$) as typical values.
Initial data for the $m$-modes of the field is calculated from the integral expressions
\begin{equation}
\psi_{m}(t=0, r) = \frac{1}{2 \pi} \int_{0}^{2 \pi} e^{- i m \phi} \psi_0(r, \phi) d\phi.
\end{equation}

\subsubsection{Method of lines and boundary conditions}
We evolved the hyperbolic partial differential equation for $\psi_m(t,r)$, Eq.~(\ref{eq:1+1}), by using the Method of Lines with fourth-order finite differencing on a uniformly-spaced grid in $r$. We used a grid with an exterior boundary condition $\psi_m(r_\text{max}) = 0$, with $r_\text{max}$ sufficiently large that the boundary is outside of causal contact with the initial pulse for the duration of the simulation. The inner boundary of the grid was placed inside the sonic horizon, at $r \approx 0.8 D$. Here, we imposed a free boundary condition, by using one-sided finite-difference stencils in the vicinity of the grid boundary. The justification for a free boundary condition is that, inside the horizon, all characteristics (i.e. all null geodesics) are inward-pointing, and thus free perturbations will naturally leave the domain. Our implementation used Mathematica and the NDSolve function.

\subsubsection{Mode sum reconstruction}
In principle, the mode sum (\ref{eq:mmode}) is an infinite series; in practice, the high-$m$ modes far above $m \sim 2\pi / \Delta \phi$ are suppressed, where $\Delta \phi$ is the minimum physical angular scale in the field. We may therefore truncate the sum. To do so without risking the introduction of high-frequency spurious features, we introduced a large-$m$ smoothing factor 
$F(m) = \frac{1}{2}\left( 1 - \tanh\left( \frac{m - m_\text{cut}}{2 \Delta m} \right) \right)$ and took the finite sum
\beq
\psi(t,r) = \text{Re} \sum_{m = 0}^{m_\text{cut} + N \Delta m} a_m F(m) \frac{\psi_m(t,r)}{\sqrt{r}} e^{i m \phi} , \quad \quad a_m \equiv \begin{cases} 1, & m=0, \\ 2, & m \neq 0. \end{cases}
\eeq
With the choice of initial data above we use $m_\text{cut} = 50$, $\Delta m = 1$ and $N = 5$.

\section{Results\label{sec:results}}
Here we present sample results for a wave sourced by an initial `pulse' disturbance in a draining bathtub flow. We examine a flow with circulation and draining parameters $C = D = 1$ subject to a `pulse' of width $\sigma = 1$ originating at $(-10, 0)$ (Sec.~\ref{sec:initialdata}). The effective horizon is at $r_h=1$, the ergosphere at $r_e = \sqrt{2} \approx 1.414$ and the co- and counter-rotating null orbits are at $r_c^+ \approx 1.082 $ and $r_c^- \approx 2.613$, respectively (see Eq.~(\ref{eq:rc})). More comprehensive results will be presented in a forthcoming work \cite{Dempsey}.

\begin{figure}
 \centering
       \includegraphics[scale=0.8]{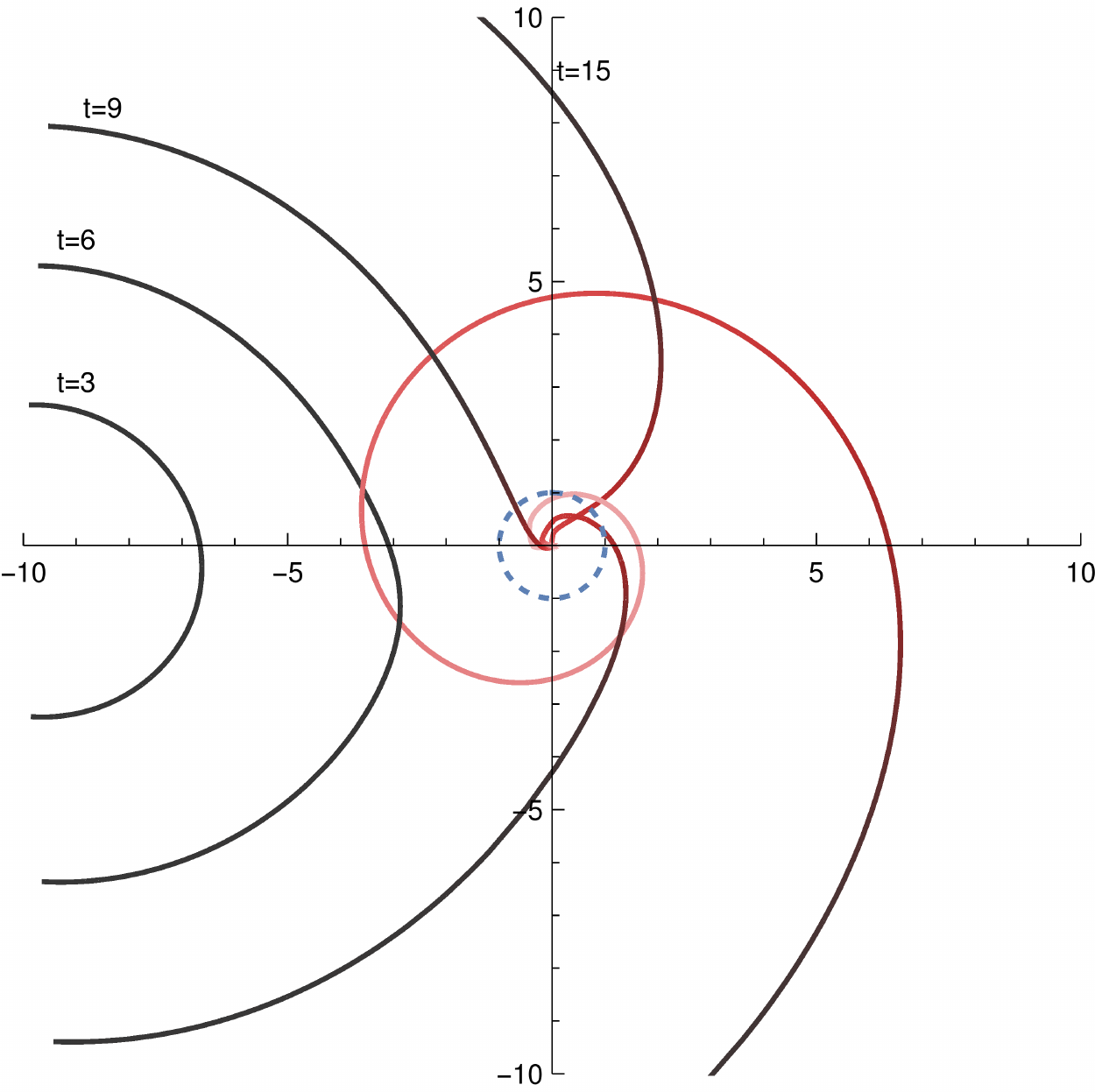}
 \caption{The plots show the evolution of a wavefront [solid line] on a draining bathtub flow with $C = 1 = D$, according to the eikonal approximation (Sec.~\ref{sec:eikonal}).  The effective horizon [dashed line] is shown at $r=1$. Here, the wavefront emanates from initial point (-10,0) at $t=0$, and is shown at time intervals $t/D = \{3,6,9,15\}$. The shading indicates the value of $\Delta^{1/2}$ on the wavefront, which is related to the amplitude $\mathcal{A}$ by Eq.~(\ref{eq:Avv}). The amplitude is weaker on the segment of the wavefront that is rapidly `stretched out' in the vicinity of the photon orbits at $r=r_c^\pm$ [Eq.~(\ref{eq:rc})]. }
 \label{fig:vvd1c1}
\end{figure}

Figure \ref{fig:vvd1c1} shows the wavefront according to the eikonal approximation (\ref{sec:eikonal}). At very early times, the wavefront emanating from $(x_0, y_0)$ is nearly circular. As time progresses, the wavefront becomes distorted. Viewed from the laboratory perspective, the co-rotating part of the wavefront (below the axis) appears to move towards the effective horizon [dashed line] more rapidly than the counter-rotating part (above the axis). Part of the wavefront falls through the effective horizon, and onto the sink point (i.e.~the `singularity') at the origin. The ergosphere (at $r_e \approx 1.414$), containing the co-rotating null orbit (at $r^+_c \approx 1.082$), causes the wavefront to `wind around' the sink. As the wavefront is stretched, it diminishes in amplitude. Near a circular null orbit, we expect the van Vleck determinant  $\Delta$ -- which determines the eikonal amplitude via Eq.~(\ref{eq:Avv}) -- to be exponentially damped with some positive Lyapunov exponent; the plot suggests this expectation is well-founded. Eventually, the wavefront intersects itself, and the points of intersection propagate away from the centre. In the non-circulating case ($C=0$), the intersections occur along the $x$ axis; in the circulating case, the points of intersection are dragged around by the flow.

The eikonal wavefront may be interpreted as a cross section of the `light-cone' on the effective geometry. In relativistic terminology, each eikonal wavefront is the intersection of a timelike hypersurface of constant $t$ with a null hypersurface (`light-cone') with its base at $(x_0, y_0)$. This suggests the possibility that, in a simple experiment, one could map out the light-cone of an effective geometry. With this motivation in mind, let us now examine whether the eikonal approximation actually provides a reliable description of a numerical solution of the wave equation (\ref{eq:wave}). 

\begin{figure}
 \centering
Propagation of a Gaussian pulse on a draining bathtub flow with circulation and draining rates $C = D = 1$.\par\medskip
       \includegraphics[scale=0.49]{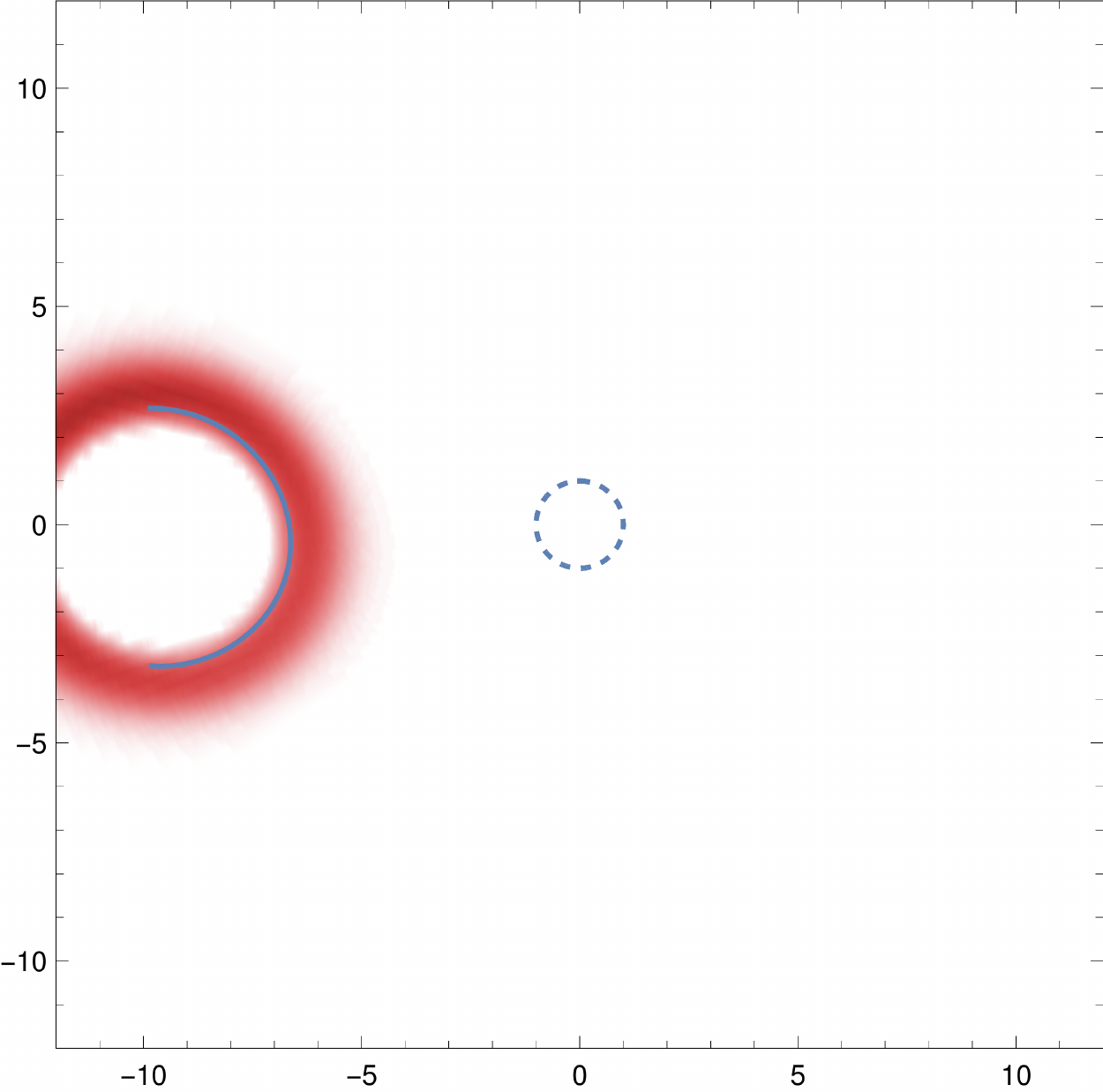}
       \includegraphics[scale=0.49]{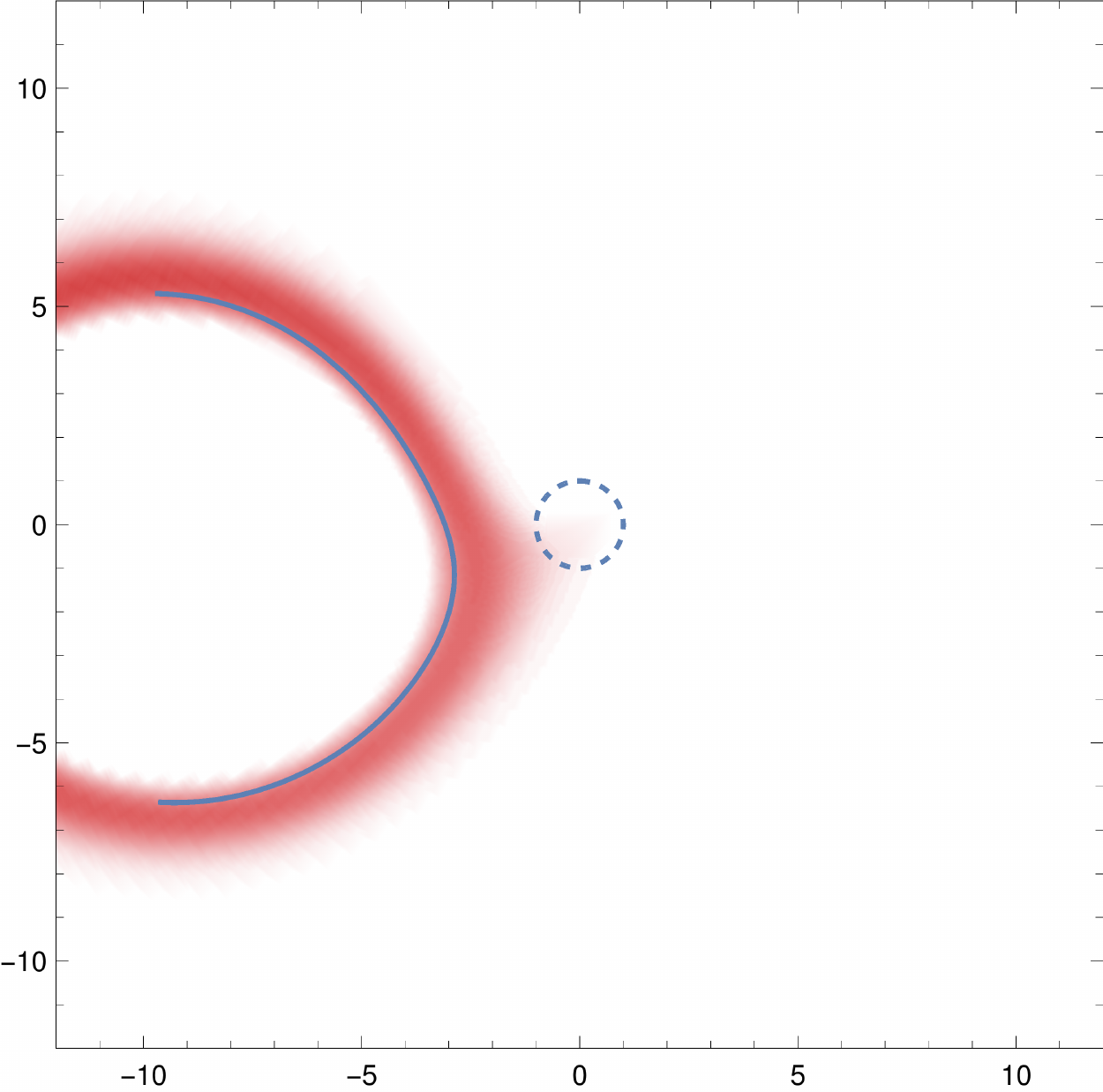}
       \includegraphics[scale=0.49]{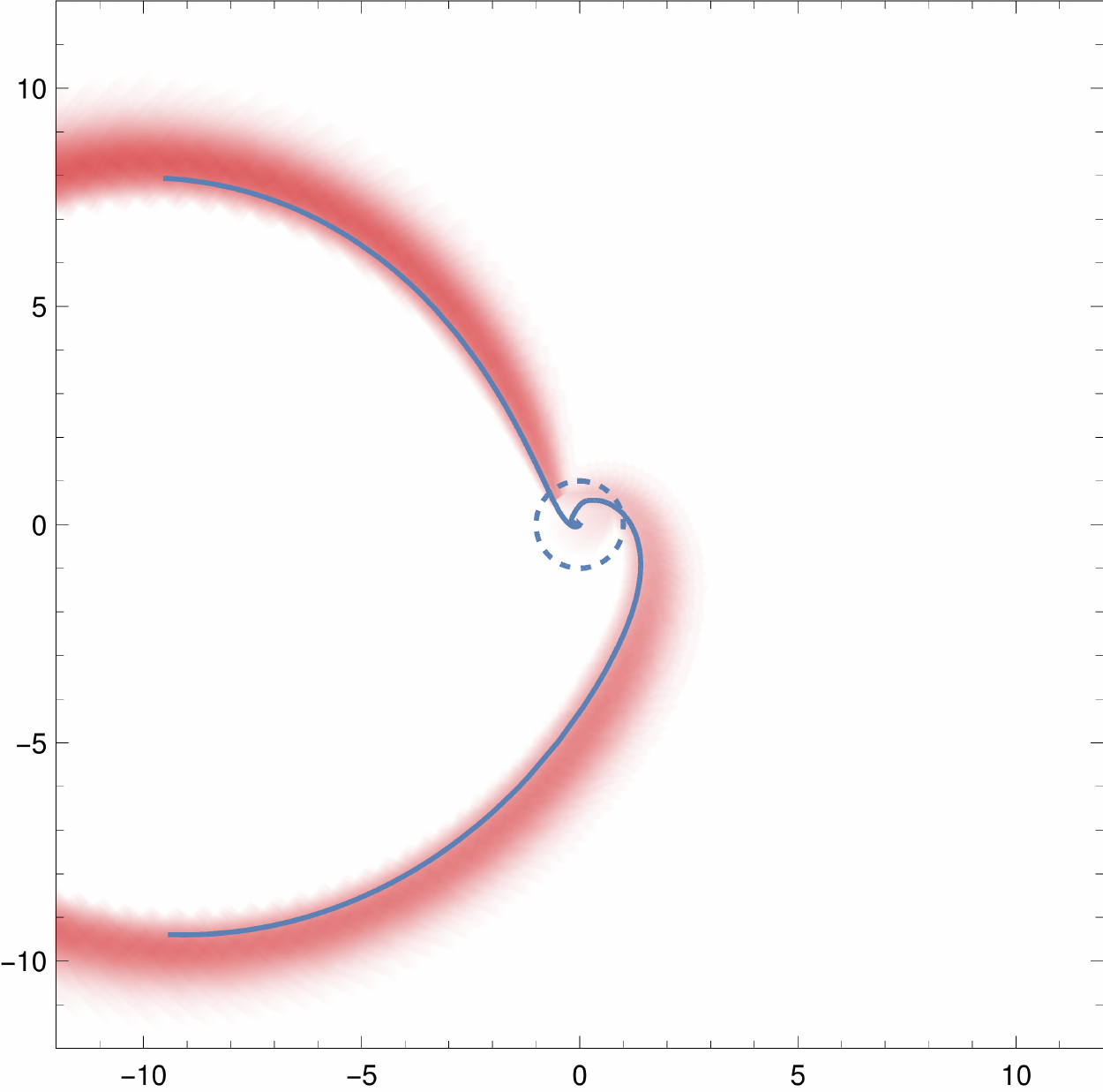}
       \includegraphics[scale=0.49]{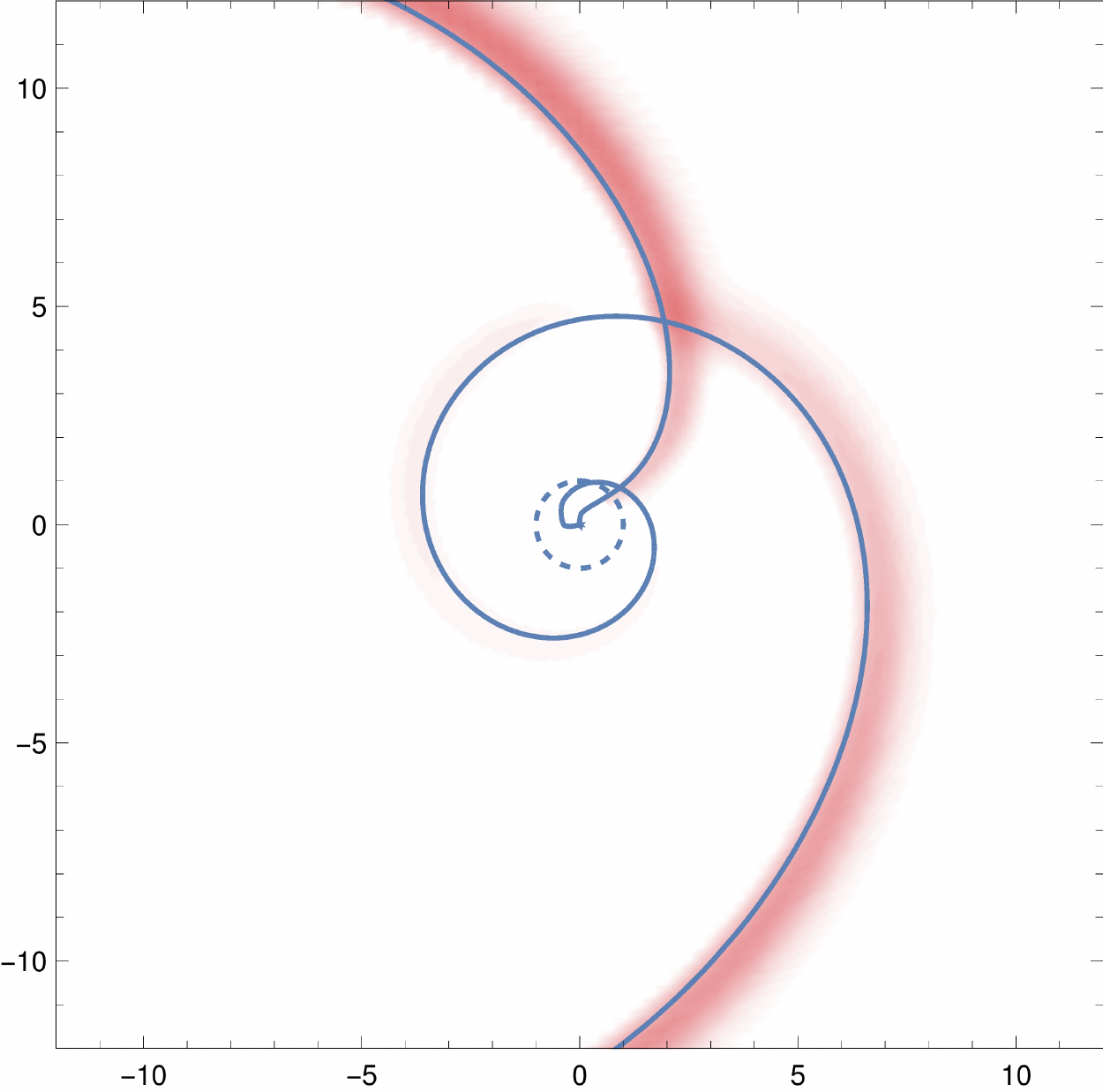}
 \caption{The plots show the evolution of a Gaussian pulse of width $\sigma = 1$ starting at (-10,0) on a draining bathtub flow with $C = 1 = D$, at times $t= \{3,6,9,15\}$. The red shading shows the amplitude of  $\psi$ found by solving wave equation (\ref{eq:wave}) using the method of Sec.~\ref{sec:waveeq}; the blue lines show the wavefront associated with $(x_0, y_0)$ according to the eikonal approximation and the transport equations (\ref{sec:eikonal})--(\ref{sec:transport}). As the disturbance propagates, it is distorted by the effective geometry, and the leading front is dragged in towards the vortex. The co-rotating part of the wavefront is dragged faster and further around the vortex. Two segments of the wavefront passing on opposite sides of the vortex intersect at a point which is dragged around in the direction of circulation. The co-rotating part of the wavefront is stretched over a longer span and thus its amplitude is diminished. The final plot ($t=15$) shows a second, weaker intersection of wavefronts.}
 \label{fig:waved1c1}
\end{figure}

Figure \ref{fig:waved1c1} shows the solution of the wave equation (red) subject to a Gaussian initial pulse (\ref{sec:initialdata}). The plots show that the propagating perturbation closely conforms to the eikonal wavefront (shown as a solid line), to a good approximation. As expected, the finite size of the pulse means that the perturbation leads the wavefront. The final plot ($t=15$) confirms that the wavefront intersects itself, as segments of the wavefront pass in opposite directions around the sink. We see a `winding' effect in the co-rotating sense caused by the ergosphere and co-rotating null orbit. At the intersection of the wavefront, it is clear that the wave amplitude is enhanced, due to constructive interference. There is also a more subtle feature: on one side of the intersection the amplitude is suppressed due to destructive interference (an initial Gaussian pulse creates a wavefront profile with a dominant maximum trailed by a sub-dominant minimum).

\section{Discussion and Conclusions\label{sec:conclusions}}
Here we have studied wave propagation on an effective curved geometry, in the context of the draining bathtub (a particular fluid-analogue model). We explored the close relationship that exists between a null geodesic congruence emanating from a point (cf. the `light-cone'), and the propagation of a `pulse' (Gaussian) disturbance governed by the wave equation on a curved geometry (\ref{eq:wave}). We have demonstrated (Fig.~\ref{fig:waved1c1}) that the propagation of such a perturbation reveals a host of interesting features of the light-cone of an analogue black hole geometry. In principle, the light-cones of black hole analogues could be studied in a fluid-mechanical experiment in the laboratory (but see caveats below).

What, if anything, can we infer about wave propagation on `astrophysical' black hole spacetimes from this draining bathtub model? Like a Kerr black hole, the draining bathtub has an (effective) horizon, an ergosphere and co- and counter-rotating null orbits. Furthermore, the eikonal approximation of Sec.~\ref{sec:eikonal} applies to either case. Thus, one should hope to be able to use our new results\cite{Dempsey} to develop intuition for how ergospheres, horizons and null orbits affect the propagation of waves. However, there is an important caveat: spacetime is 3+1 dimensional, whereas surface waves on a draining bathtub is a 2+1 dimensional model. This leads to some important differences. First, in flat spacetime ($d=3$), waves propagate according to Huygen's principle, and the retarded Green function has (distributional) support only on the light-cone, whereas in $d=2$, the Green function also has extended support within the light-cone. Second, in $d=3$ the intersection of wavefronts typically generates caustics (focal points of a one-parameter family of null rays\cite{Perlick:2004tq, Bozza:2008mi, Casals:2009zh, Dolan:2011fh, Zenginoglu:2012xe}), whereas in $d=2$ intersections lead only to amplitude `doubling'. Third, `topological' features such the Aharonov-Bohm effect are possible in $d=2$ but not $d=3$\cite{Berry:1980, Dolan:2011zza}.  

Let us now review the model with a critical eye. The first key assumption here is that all wavelengths propagate at the same speed; in other words, we have assumed a linear dispersion relation $\omega = c k$, where $\omega$ is the frequency, $k=2\pi / \lambda$ the wavenumber, and $c$ is the (constant) speed of propagation. In fact, the dispersion relation for (e.g.) gravity-capillary waves in a tank of depth $d$ is $\omega^2 = \left(g k + \sigma k^3 / \rho \right) \tanh( k d )$, where $g$ the gravitational field, and $\sigma$ is the surface tension (in N/m). While the dispersion relation is approximately linear for long wavelengths ($\omega \approx \sqrt{g d} \, k$), this assumption breaks down for wavelengths $\lambda \lesssim d$ or $\lambda \lesssim \sqrt{\sigma / \rho g}$. Thus, very narrow `pulses' may propagate quite differently from long wavelength perturbations. However, we have shown here in Fig.~\ref{fig:waved1c1} (with $\sigma = 1$) that it is not necessary to use a very narrow pulse in order to probe the effective geometry.

The second key assumption is that it is experimentally feasible to maintain the stability of the converging flow as it becomes supersonic in an ergoregion. Real bathtub vortices are rich fluid-mechanical systems\cite{PhysRevLett.91.104502}, and present many challenges for the experimentalist\cite{Berry:1980, QGL}. It may be that Eq.~(\ref{eq:wave}) is more accessible in analogues in other media, such as Bose-Einstein condensates or materials of variable refractive index. Whichever approach is taken, the prospect of understanding wave propagation around black holes through laboratory experiments surely provides a compelling motivation.

\section*{Acknowledgments}
The work of SRD is supported by the Lancaster-Manchester-Sheffield Consortium for Fundamental Physics under STFC grant ST/L000520/1, and by EPSRC under grant EP/M025802/1. 

\bibliographystyle{ws-ijmpd}
\bibliography{references}

\end{document}